\documentclass[11pt]{article}

\usepackage{amsmath,amssymb,color,graphicx,cite}
\usepackage{titlesec}
\usepackage[titletoc,title]{appendix}
\usepackage{amsthm}
\usepackage{subcaption}
\usepackage[section]{placeins}
\usepackage{float}

\newcommand{\hoch}[1]{$\, ^{#1}$}

\usepackage[normalem]{ulem}

\newcommand{\be}{\begin{equation}}
\newcommand{\ee}{\end{equation}}
\newcommand{\bea}{\setlength\arraycolsep{2pt} \begin{eqnarray}}
\newcommand{\eea}{\end{eqnarray}}
\newcommand{\nn}{\nonumber}

\def\cramp{\medmuskip = 2mu plus 1mu minus 2mu}

\def\fft#1#2{{\frac{#1}{#2}}}

\def\0{{\sst{(0)}}}
\def\1{{\sst{(1)}}}
\def\2{{\sst{(2)}}}
\def\3{{\sst{(3)}}}
\def\4{{\sst{(4)}}}
\def\5{{\sst{(5)}}}
\def\6{{\sst{(6)}}}
\def\7{{\sst{(7)}}}
\def\8{{\sst{(8)}}}
\def\sst#1{{\scriptscriptstyle #1}}

\begin{document}

\begin{center}
{\Large {\bf When Bumblebee Meets NLED: Lorentz-Violating Black Holes and Regular Spacetimes}}

\vspace{20pt}
Zhi-Chao Li\hoch{1} and H.~L\"u\hoch{1,2}

\vspace{10pt}

{\it \hoch{1}Center for Joint Quantum Studies, Department of Physics,\\
School of Science, Tianjin University, Tianjin 300350, China }

\medskip

{\it \hoch{2}The International Joint Institute of Tianjin University, Fuzhou,\\
Tianjin University, Tianjin 300350, China}

\vspace{40pt}
\end{center}

\begin{abstract}

We construct charged black hole solutions in bumblebee gravity coupled to a general class of nonlinear electrodynamics (NLED) using an auxiliary Maxwell-scalar formalism. The norm-fixed radial configuration of the bumblebee vector makes the solutions asymptotic to a conical Lorentz-violating vacuum and requires stringent nonminimal bumblebee-NLED couplings. The general black hole solutions contain independent mass and charge parameters. There are two sources of singular behavior at the center: one is due to the Schwarzschild-type pole and the other is the residual conical singularity of the Lorentz-violating vacuum. By fine-tuning the mass-charge relation, one can generally remove the pole singularity, giving rise to marginally regular black holes. For a suitable NLED theory such as Born-Infeld theory, both singularity sources can be removed at the cost of requiring both the mass and the charge to be fine-tuned to specific functions of the coupling constants. The resulting solutions describe regular horizonless spacetimes interpolating from AdS or dS cores to Lorentz-violating vacua.

 \end{abstract}

\vfill{\footnotesize lizc@tju.edu.cn \ \ \ mrhonglu@gmail.com}

\thispagestyle{empty}
\pagebreak

\setcounter{tocdepth}{2}
\tableofcontents
\pagebreak

\section{Introduction}
\label{sec:intro}

Lorentz invariance is one of the foundational symmetries of modern high-energy physics and general relativity. From the effective-field-theory viewpoint, controlled violations of Lorentz symmetry provide a systematic way to parameterize possible ultraviolet effects beyond the standard low-energy description. This idea has been extensively developed in the framework of the Standard-Model Extension and its gravitational sector \cite{KosteleckySamuel:1989,KosteleckyRussell:2008,Mattingly:2005,BaileyKostelecky:2006}. Among the simplest generally covariant realizations are bumblebee models, in which a vector field acquires a nonzero vacuum expectation value and thereby selects a preferred spacetime direction. In this way local Lorentz symmetry is broken spontaneously while the theory remains formulated in covariant language \cite{BluhmKostelecky:2005}.

Bumblebee gravity provides a useful arena for studying how Lorentz violation modifies gravitational dynamics and black hole geometries. Static and spherically symmetric Schwarzschild-like solutions have been constructed in bumblebee gravity \cite{Casana:2017jkc}. A characteristic feature is that the norm of the bumblebee vector is a constant throughout the spacetime. Consequently, the geometry is not asymptotic to the usual Minkowski vacuum, but to a conical Lorentz-violating vacuum. This is qualitatively different from ordinary Maxwell or Proca hair, whose field strengths may decay at large radius without leaving a fixed spacelike vector expectation value \cite{Geng:2015kvs,Fan:2016jnz,Babichev:2017rti,Heisenberg:2017xda,Xu:2023}.

This conical Lorentz-violating background raises a sharper regularity question than in ordinary asymptotically flat black hole spacetimes.  Now, there are two distinct sources of singular behavior: the usual Schwarzschild-type mass singularity and the conical central defect associated with the Lorentz-violating vacuum itself. Removing the Schwarzschild-type pole is therefore not enough to obtain a regular center. It is then natural to ask whether the standard mechanisms for constructing regular black holes can be implemented consistently in a Lorentz-violating bumblebee background.

Regular black holes in Einstein gravity have been studied for a long time. The Bardeen model and subsequent developments showed that the central Schwarzschild singularity may be replaced by a regular core \cite{Bardeen:1968,Dymnikova:1992,Ansoldi:2008, Lan:2023cvz}, often with a de Sitter, anti-de Sitter or even Minkowski \cite{Xiang:2013sza} core. In the spherically-symmetric and static configurations, the latter two cases will necessarily violate the null energy condition \cite{Li:2023yyw}. At the early stage, the construction of regular black holes, such as the Bardeen black hole, was purely metric-based, with the required minimally coupled matter energy-momentum tensor satisfying some suitable energy conditions. However, this approach does not work well in bumblebee gravity.  As we shall see later, minimally coupled matter is not compatible with the norm-fixed ansatz of the bumblebee vector for the Lorentz-violating spacetime. We thus must specify a fundamental matter Lagrangian.

NLED provides one of the standard matter mechanisms. NLED was independently motivated from field theory. Born-Infeld theory was introduced to soften the divergence of the Maxwell self-energy \cite{BornInfeld:1934}, and Born-Infeld-type structures arise naturally in low-energy effective actions of string theory \cite{FradkinTseytlin:1985}. Suitable NLED sectors can support nonsingular or nearly nonsingular charged geometries in Einstein gravity \cite{AyonBeatoGarcia:1998,Bronnikov:2001}. It turns out a reverse-engineering technique allows one to construct a corresponding NLED with magnetic charges for all spherically-symmetric and static spacetimes with constant $g_{tt} g_{rr}$ \cite{Fan:2016hvf}. More generally, NLED theories provide a flexible framework for studying modified charged black holes, electromagnetic duality, and strong-field corrections beyond Maxwell theory \cite{GibbonsRasheed:1995}. Nevertheless, constructing regular electric black holes directly from a prescribed single-valued Lagrangian of the type $L(\mathcal F)$ with ${\cal F}=F^{\mu\nu}F_{\mu\nu}$ is prohibited by no-go theorems \cite{AyonBeatoGarcia:1998,Bronnikov:2001,Li:2024rbw}.

A useful way to control this branch structure is the Einstein-Maxwell-scalar (EMS) theory. In this formulation the gauge field remains bilinear in $F_{\mu\nu}$, while the nonlinear electromagnetic response is encoded in an auxiliary scalar without a kinetic term. Eliminating this scalar on a chosen algebraic branch reproduces an effective NLED theory. This representation is particularly convenient for constructing analytic charged geometries and for regular electric black hole models \cite{Li:2023yyw,Li:2024rbw}.

The central difficulty in the present problem is that ordinary minimally coupled matter is not generically compatible with the norm-fixed radial bumblebee vector ansatz. The Einstein equations and the bumblebee-vector equation simultaneously constrain the metric, the vector profile, and the matter sector. Even for the linear Maxwell field, constructing charged bumblebee black holes requires carefully chosen nonminimal couplings between the bumblebee vector and the electromagnetic sector \cite{Liu:2024charged,Li:2025tcd, Chen:2025ypx,Ovcharenko:2026rvj}. A direct coupling of a generic nonlinear function $L(\mathcal F)$ to the bumblebee background would therefore be technically cumbersome and may not lead to a consistent radial system. The auxiliary Maxwell-scalar representation of NLED offers a more tractable route, because the nonlinear electromagnetic response is controlled algebraically while the gauge-field dependence remains bilinear.

The rest of this paper is organized as follows. In Sec.~\ref{sec:nled-bumblebee}, we formulate the bumblebee-deformed Einstein-NLED theory using the auxiliary Maxwell-scalar representation and derive the reduced master equations for static and spherically symmetric configurations. In Sec.~\ref{sec:regular-bh}, we construct explicit electrically charged solutions, including the Born-Infeld case and a one-parameter EMS family, and discuss the thermodynamics of the unconstrained black hole solutions. In Sec.~\ref{sec:central_regular_horizon}, we analyze the central regularity and horizon structure, showing how marginally regular black holes and regular horizonless geometries arise in different solutions. Finally, Sec.~\ref{sec:con} summarizes the results and discusses possible extensions.

\section{Bumblebee gravity and NLED matter}
\label{sec:nled-bumblebee}

\subsection{Bumblebee gravity and matter coupling}

We begin with a brief review of bumblebee gravity. The minimal field content consists of the metric and a vector field (i.e.~1-form potential $B_\1=B_\mu dx^\mu$) that is nonminimally coupled to gravity.
The Lagrangian takes the form
\be
{\cal L}_{\rm B}
=
\sqrt{-g}
\Big(
R
+\xi\,B^\mu B^\nu R_{\mu\nu}
-\frac14 B_{\mu\nu}B^{\mu\nu}
-U(B^2-b^2)
\Big)\,,
\label{bumblebeegrav}
\ee
where $\xi$ is the nonminimal coupling constant, $B^2=B^\mu B_\mu$, and
$B_{\mu\nu}=\partial_\mu B_\nu-\partial_\nu B_\mu$ is the field strength of
the vector field $B$. The theory was constructed to describe a vacuum in which local Lorentz symmetry is spontaneously broken. More precisely, a nonzero vacuum expectation value of $B_\mu$ selects a preferred local spacetime direction and therefore breaks local Lorentz symmetry spontaneously. This should be distinguished from ordinary Maxwell or Proca hair, whose field strength can decay at large radius. For example, electric-type vector profiles typically decay at large radius so that the asymptotic region may approach a Lorentz-invariant or asymptotically flat vacuum, as happens in several vector-tensor black hole constructions \cite{Geng:2015kvs,Fan:2016jnz, Babichev:2017rti,Heisenberg:2017xda}.

In order to violate Lorentz symmetry in the vacuum, one approach is to lock the vector field $B_\1$ to lie only in the radial direction so that its norm is fixed ($B^2=b^2$) for some constant $b$, by introducing a suitable potential $U$. For this norm-fixed radial bumblebee vector, the nonminimal curvature coupling in \eqref{bumblebeegrav} admits a static and spherically symmetric Lorentz-violating vacuum. The corresponding vacuum geometry is simply a direct product of time and a three-dimensional Euclidean cone, namely
\be
ds^2= -dt^2 + (1+\ell)\,dr^2 + r^2d\Omega_2^2\,, \qquad B_\1=b \sqrt{1+\ell}\, dr\,,
\qquad \ell=\xi b^2\,.\label{LVvac}
\ee
The Euclidean cone has a power-law curvature singularity at $r=0$, with the Ricci scalar $R=2\ell/[(1+\ell)r^2]$. In the following we assume $1+\ell>0$, so that the radial metric component has the correct signature. It turns out that the theory also admits a black hole solution that is asymptotic to this vacuum cone, i.e. \cite{Casana:2017jkc}
\begin{equation}
ds^2= -(1+\ell)f(r) dt^2+ \frac{dr^2}{f(r)} +r^2 d\Omega_2^2,
\qquad f(r)=\frac{1}{1+\ell}-\frac{2m}{r}, \qquad B_\1=\frac{b}{\sqrt{f(r)}}\,dr .
\label{eq:bumblebee-bh}
\end{equation}
Here $m$ is an integration constant, while $b$ is a
global parameter of the theory.
Because this solution differs from the Schwarzschild geometry mainly through
the conical Lorentz-violating deformation, it provides a useful starting point
for studying matter deformations of bumblebee black holes.

A natural question arises: how does the matter energy-momentum tensor affect
the black hole \eqref{eq:bumblebee-bh}? This turns out to be quite subtle. To
see this, we consider a Lagrangian with bumblebee gravity and additional minimally coupled
matter, namely
\be
{\cal L}
=
{\cal L}_{\rm B}
+
{\cal L}_{\rm matter}\,.
\ee
Then the Einstein equation becomes
\begin{align}
G_{\mu\nu}
&=
T^{\rm matter}_{\mu\nu}
-
\xi
\Big[
2R_{\rho(\mu}B_{\nu)}B^\rho
-
\frac12 R_{\rho\sigma}B^\rho B^\sigma g_{\mu\nu}
-
\nabla_\rho\nabla_{(\mu}\left(B_{\nu)}B^\rho\right)
\nonumber\\
&\hspace{3.0cm}
+
\frac12 \Box\left(B_\mu B_\nu\right)
+
\frac12 g_{\mu\nu}
\nabla_\rho\nabla_\sigma
\left(B^\rho B^\sigma\right)
\Big]
+
\cdots ,
\label{eq:einstein-bumblebee}
\end{align}
where we present only the relevant part for discussion.
On the vacuum branch $U_Y=0$, the equation associated with the variation of $B_\mu$ reduces to
\be
\nabla_\mu B^{\mu\nu}
=
-2\xi R^\nu{}_{\mu}B^\mu\,.
\label{eq:B-eom}
\ee

For the norm-fixed radial ansatz $B_\mu dx^\mu\sim b\,dr$, the Einstein equation \eqref{eq:einstein-bumblebee} and the $B_\mu$ equation \eqref{eq:B-eom} generally impose more algebraic and differential constraints
than can be satisfied by a generic minimally coupled matter stress tensor.
Thus, within this restricted ansatz, minimally coupled matter does not in
general lead to a simple matter-deformed version of the bumblebee black hole
\eqref{eq:bumblebee-bh}. In this restricted sense, the ansatz exhibits an
effective no-minimal-matter-hair property. However, it should be emphasized that equations of motion from a Lagrangian are necessarily consistent. When we say the two equations
\eqref{eq:einstein-bumblebee} and \eqref{eq:B-eom} are inconsistent, we mean only in the context of the specific ansatz of the $B$ field. This should not be interpreted as a general no-hair theorem for all bumblebee configurations.

In order to resolve this over constraint associated with the norm-fixed radial bumblebee vector, one is naturally led to include nonminimal couplings between the matter sector and the bumblebee vector field. This effort was pioneered in \cite{Liu:2024charged}, where an electromagnetic field was nonminimally coupled to the bumblebee vector and exact charged spherically symmetric, as well as slowly rotating black hole solutions were constructed with and without a cosmological constant. Further subtleties in black hole thermodynamics were later resolved in \cite{Chen:2025ypx,Li:2025tcd}.

The present work extends this idea from linear Maxwell electrodynamics to
a class of NLED. Instead of introducing a prescribed nonlinear
function $L(\mathcal F)$ directly, we use an auxiliary Maxwell-scalar
representation in which the nonlinear constitutive relation is encoded by two
functions $Z(\phi)$ and $V(\phi)$. This formulation is particularly useful
for the bumblebee problem, because the nonminimal couplings between $B_\mu$
and the electromagnetic field deform the algebraic scalar equation rather than
turning it into a differential equation. As a result, the matter sector can
remain analytically tractable while still capturing NLED effects.

\subsection{NLED as Maxwell-scalar theory}
\label{subsec:nled-as-ems}

There are several ways to generalize Maxwell theory to include
nonlinear electromagnetic corrections. In this paper, we focus on a simple class in which the Maxwell kinetic term is promoted to a generic function of the standard electromagnetic invariant,
\be
L=L({\cal F})\,,
\qquad
\hbox{with}
\qquad
{\cal F}
=
\frac{1}{4} F^{\mu\nu}F_{\mu\nu}\,,
\qquad
F_{\mu\nu}
=
\partial_\mu A_\nu-\partial_\nu A_\mu\,.
\label{NLEDlag}
\ee
Thus the theory reduces to Maxwell theory when $L=-{\cal F}$.
We further require that in the weak field limit, it behaves like \cite{Li:2023yyw,Li:2024rbw}
\be
L=-{\cal F} + \alpha_1{\cal F}^2 + \alpha_2{\cal F}^3 +\cdots\,.
\ee
In other words, in the weak field limit, $L$ is an analytic function of
${\cal F}$ around the Maxwell point. This property allows one to view the
theory as an effective theory of a possible ultraviolet completion. Note that this analyticity requirement rules out the possibility of constructing a large class of regular black holes, including the Bardeen or Hayward black holes, within the Einstein-NLED theory.

It was proposed that the NLED of \eqref{NLEDlag} can be equivalently described by a Maxwell-scalar (MS) theory where the scalar is auxiliary,
\begin{equation}
L=-Z(\phi)\,\mathcal F-V(\phi)\,.
\label{eq:EMS_auxiliary}
\end{equation}
Here $\phi$ carries no kinetic term and is therefore an auxiliary field.
Varying with respect to $\phi$ gives
\begin{equation}
V'(\phi)+Z'(\phi)\,\mathcal F=0 .
\label{eq:phi_auxiliary_general}
\end{equation}
This is an algebraic equation for the auxiliary scalar $\phi$.
On a given branch, it determines
\begin{equation}
\phi=\phi({\cal F}) .
\end{equation}
Substituting this solution back into \eqref{eq:EMS_auxiliary}, one recovers an
effective NLED Lagrangian
\begin{equation}
L_{\rm eff}(\mathcal F)
=
-
Z\bigl(\phi(\mathcal F)\bigr)\,\mathcal F
-
V\bigl(\phi(\mathcal F)\bigr).
\label{eq:NLED_from_auxiliary}
\end{equation}

There is a caveat to the claim that the MS theory and the NLED $L({\cal F})$
are equivalent. The algebraic scalar equation
\eqref{eq:phi_auxiliary_general} can generally have multiple branches.
Therefore, a single Maxwell-scalar theory may generate several effective NLED
Lagrangians after $\phi$ is eliminated, depending on which branch is chosen.
Conversely, a prescribed NLED $L({\cal F})$ may be reproduced only locally in
field space. This branch structure is especially useful in the construction of
regular electric black holes, where purely electric regular solutions in
Einstein-NLED theory are constrained by known no-go results
\cite{AyonBeatoGarcia:1998,Bronnikov:2001,Li:2024rbw}.
In particular, regular electric black holes in Einstein-NLED gravity
typically require a multi-branched or non-single-valued effective NLED
description when written directly as $L({\cal F})$
\cite{AyonBeatoGarcia:1998,Bronnikov:2001}.
This obstruction can be circumvented more naturally in the auxiliary
Maxwell-scalar formulation, because the auxiliary scalar equation can select
different algebraic branches while the underlying Lagrangian remains
single-valued \cite{Li:2024rbw}.

One goal of this paper is to construct bumblebee-deformed Einstein-NLED
theories that admit Lorentz-violating black holes. As we have pointed out earlier, the nonminimal coupling between the bumblebee vector $B_\mu$ and the matter field is not only necessary, but
also very tricky.
In the Maxwell case, suitable nonminimal couplings between
$B_\mu$ and the gauge sector have been used to construct charged bumblebee
black holes \cite{Li:2025tcd,Liu:2024charged}. In the literature, how the Maxwell field should be coupled to $B_\mu$
has now been well understood, and we find that it is much simpler to find a
suitable coupling of the Maxwell-scalar sector to $B_\mu$.

Since the scalar in \eqref{eq:EMS_auxiliary} is auxiliary, the term
Maxwell-scalar theory should not be understood as introducing a propagating
scalar degree of freedom. In what follows, we shall also refer to this
auxiliary Maxwell-scalar representation simply as an NLED theory.

\subsection{Bumblebee-deformed Einstein-NLED gravity}
\label{subsec:bumblebee-ems}

In this subsection, we present the bumblebee-deformed
Einstein-NLED theory, where the NLED sector is written in the auxiliary
Maxwell-scalar form \eqref{eq:EMS_auxiliary}.
Our goal is to construct Lorentz-violating, static and spherically symmetric
spacetime geometries sourced by nonlinear electromagnetic fields. At the level
of the reduced equations, we allow both electric and magnetic parameters,
although the explicit regularity analysis below will be restricted to the
purely electric branch.
We consider the following theory
\begin{align}
{\cal L} = \sqrt{-g}\,\Big[
&R- Z(\phi)\,\mathcal F
- V(\phi)
-\frac14 B_{\mu\nu}B^{\mu\nu}
+\xi\,B^\mu B^\nu R_{\mu\nu}
+\xi_1\,B^2\,V(\phi)
-U(Y) \nonumber\\
&\quad
+ Z(\phi)\big(\gamma_1 + \gamma_3 B^2\big)\,\big(B_\mu F^{\mu\nu}\big)\big(B^\rho F_{\rho\nu}\big)
+ Z(\phi)\big(\gamma_2 + \gamma_4 B^2\big)\,B^2\,\mathcal F
\Big],
\label{eq:action}
\end{align}
where $Y\equiv B^2- b^2$.
The first three terms in the square bracket describe Einstein-NLED gravity in the auxiliary Maxwell-scalar representation. The remaining terms encode the
bumblebee deformation. In addition to the original bumblebee parameters
$(\xi,b)$, we introduce five matter-coupling constants
$(\xi_1,\gamma_1,\gamma_2,\gamma_3,\gamma_4)$. The electromagnetic field
strength is kept bilinear throughout the Lagrangian, while its couplings to
the bumblebee vector are retained up to and including the $B^4$ order. As we shall see later, this gives us more parameters than are needed for constructing Lorentz-violating black holes. However, the over-parametrization includes examples that may allow a novel generating technique proposed in \cite{Ovcharenko:2026rvj}.

The matter couplings in \eqref{eq:action} are chosen to provide a general parity-even parametrization that is bilinear in $F_{\mu\nu}$, algebraic in the auxiliary scalar, and contains powers of $B^2$ up to $B^4$ within the static spherical branch considered here. Parity-odd couplings involving $F_{\mu\nu}\widetilde F^{\mu\nu}$ are not included, since the explicit regularity analysis below is restricted to the purely electric sector.

We now present the covariant equations of motion associated with the variations
of all the fields in the theory. Varying the Lagrangian \eqref{eq:action} with
respect to $A_\mu$, $\phi$, and $B_\mu$ yields
\bea
\delta A_\mu: && 0=\nabla_\mu {\cal P}^{\mu\nu}\,,\quad
{\cal P}^{\mu\nu} = Z(\phi)\left[\big((\gamma_2+\gamma_4B^2)B^2-1\big)F^{\mu\nu}
+4\big(\gamma_1+\gamma_3B^2\big)B^{[\mu}X^{\nu]}\right],
\label{eq:A_general}
\\[2mm]
\delta \phi: && 0=(\xi_1 B^2-1)V'(\phi)
+Z'(\phi)\Big[\big(\gamma_1+\gamma_3 B^2\big)X_\mu X^\mu
+\big((\gamma_2+\gamma_4 B^2)B^2-1\big)\mathcal F\Big],
\label{eq:phi_general}
\\[2mm]
\delta B_\mu: && 0=\nabla_\mu B^{\mu\nu}
+2\xi R^\nu{}_{\mu}B^\mu
+2\xi_1V(\phi)B^\nu
-2U_Y B^\nu
+2Z(\phi)\big(\gamma_1+\gamma_3B^2\big)F^{\nu\rho}X_\rho
\nonumber\\
&&\hspace{2.6cm}
+2Z(\phi)\Big[\gamma_3 X_\rho X^\rho+\big(\gamma_2+2\gamma_4 B^2\big)\mathcal F\Big]B^\nu,
\label{eq:B_general}
\eea
where, for notational convenience, we define
$X_\mu\equiv B^\nu F_{\nu\mu}$ and
$U_Y\equiv dU/dY$. The variation of the metric gives rise to the Einstein equation
$\mathcal E_{\mu\nu}=0$, where
\cramp{
\begin{align}
\mathcal E_{\mu\nu}&\equiv\;
R_{\mu\nu}-\frac12  g_{\mu\nu} R
-\frac12 Z(\phi)\,F_{\mu\rho}F_{\nu}{}^{\rho}
+\frac12 Z(\phi)\,\mathcal F\, g_{\mu\nu}
+\frac12\bigl(V(\phi)+U(Y)\bigr)g_{\mu\nu}
\nonumber\\
&+\xi_1 V(\phi)\Big(B_\mu B_\nu-\frac12 B^2 g_{\mu\nu}\Big)
-U_Y B_\mu B_\nu
-\frac12 B_{\mu\rho}B_{\nu}{}^{\rho}
+\frac18 B_{\rho\sigma}B^{\rho\sigma}g_{\mu\nu}
\nonumber\\
&+\xi\Big[
2R_{\rho(\mu}B_{\nu)}B^\rho
-\frac{1}{2} R_{\rho\sigma}B^\rho B^\sigma g_{\mu\nu}
-\nabla_\rho\nabla_{(\mu}\!\big(B_{\nu)}B^\rho\big)
+\frac12 \Box(B_\mu B_\nu)
+\frac12 g_{\mu\nu}\nabla_\rho\nabla_\sigma(B^\rho B^\sigma)
\Big]
\nonumber\\
&+Z(\phi)\bigg[
(\gamma_1+\gamma_3 B^2)\Big(
F_{\mu\rho}F_{\nu\sigma}B^\rho B^\sigma
+2 B_{(\mu}F_{\nu)}{}^{\rho}X_\rho
-\frac12 g_{\mu\nu}X_\rho X^\rho
\Big)
+\gamma_3\,X_\rho X^\rho\,B_\mu B_\nu \nonumber\\
& + \frac12(\gamma_2+\gamma_4 B^2)\,B^2\,F_{\mu\rho}F_{\nu}{}^{\rho}
+(\gamma_2+2\gamma_4 B^2)\,\mathcal F\, B_\mu B_\nu
-\frac12(\gamma_2+\gamma_4 B^2)\,B^2\,\mathcal F\, g_{\mu\nu}
\bigg].
\label{eq:Einstein_general}
\end{align}}

In this paper, we consider spherically symmetric and static spacetime geometries. The most general ansatz for the metric is
\begin{equation}
ds^2=-h(r)\,dt^2+\frac{dr^2}{f(r)}+r^2(d\theta^2+\sin^2\theta\,d\varphi^2).
\label{eq:metric_ansatz}
\end{equation}
The symmetry-breaking potential $U(Y)$ enforces a nonvanishing vacuum
expectation value for $B_\mu$, so that on the vacuum branch one has
$B_\mu B^\mu= b^2$, ensured by $U(0)=0=U_Y(0)$. For the bumblebee vector, we take
\begin{equation}
B_\mu dx^\mu = B_r(r)\,dr, \qquad B_r(r)=\frac{b}{\sqrt{f(r)}}\,.
\label{eq:B_ansatz}
\end{equation}
This branch satisfies $B^2=b^2$ identically, and we therefore refer to it as the norm-fixed radial bumblebee vector.

The Maxwell field can carry both electric and magnetic charges.
We use the gauge potential
\begin{equation}
A_\mu dx^\mu = a(r)\,dt + 4P_m\cos\theta\, d\varphi,
\qquad
F_{tr}=-a'(r),\qquad
F_{\theta\varphi}=-4P_m\sin\theta\,.
\label{eq:A_ansatz}
\end{equation}
Here $P_m$ is the magnetic charge parameter. It follows from the generalized
Maxwell equation \eqref{eq:A_general} that $a'$ can be solved as
\begin{equation}
a'(r)=\frac{4q\sqrt{h(r)}}{r^2\sqrt{f(r)}\,Z(\phi)}\,,
\label{eq:a_first_integral}
\end{equation}
where $q$ is an electric charge parameter.

We therefore have
\be
\mathcal F
=-\fft{8q^2}{r^4 Z(\phi)^2}+\frac{8P_m^2}{r^4}\,,\qquad
X_\mu X^\mu =-\fft{16 b^2 q^2}{r^4 Z(\phi)^2}\,.
\label{eq:X2_invariant}
\ee
The scalar equation \eqref{eq:phi_general} then reduces to
\begin{equation}
(\xi_1 b^2-1)V'(\phi)
+\frac{8}{r^4}
\left(\frac{\Delta_e q^2}{Z(\phi)^2}-P_m^2\Delta_m\right)Z'(\phi)=0\,,
\label{eq:phi_master}
\end{equation}
where we have introduced two charge-renormalization parameters $\Delta_e$ and
$\Delta_m$, specified by the bumblebee deformation parameters:
\begin{equation}
\Delta_e\equiv 1-b^2(2\gamma_1+\gamma_2)-b^4(2\gamma_3+\gamma_4),
\qquad
\Delta_m\equiv 1-b^2\gamma_2-b^4\gamma_4.
\label{eq:Delta_defs}
\end{equation}
They determine how the nonminimal bumblebee-electromagnetic couplings rescale
the electric and magnetic source terms in the reduced radial equations.
Next, the radial component of the $B_\mu$ equation
\eqref{eq:B_general} contains $h''(r)$ linearly, which can be solved in terms of
$h$, $h'$, $f$, $f'$, $\phi$ and the matter profiles. Substituting the resulting expression
into the Einstein equations, we find that the combination
$-\mathcal E^{t}{}_{t}+\mathcal E^{r}{}_{r}=0$ collapses to the simple
first-order relation
\begin{equation}
(1+b^2\xi)\,\frac{h(r)f'(r)-f(r)h'(r)}{r\,h(r)}=0.
\label{eq:hf_combination}
\end{equation}
Therefore, for generic $1+b^2\xi\neq 0$, we have $h(r)=c f(r)$ for a
constant $c$. Consequently, we now have
\be
a'(r)=\frac{4\sqrt{c}\, q}{r^2 Z(\phi)}\,.
\label{eq:a_first_integral_hf}
\ee
The $tt$ component of the Einstein equation, $\mathcal E^{t}{}_{t}=0$, can
now be cast into a first-order flow equation for $f$, namely
\begin{equation}
r f'(r)+f(r)-\frac{1}{1+b^2\xi}
= -\frac{r^2(1-b^2 \xi_1)}{2(1+b^2\xi)}\,V(\phi)
-\frac{4q^2\Delta_e}{r^2(1+b^2\xi)\,Z(\phi)}
-\frac{4P_m^2\Delta_m\,Z(\phi)}{r^2(1+b^2\xi)} .
\label{eq:f_master_general}
\end{equation}
At this stage, the remaining Einstein equations, as well as the radial
bumblebee equation, must be compatible with the reduced equations above,
because all functions in the ansatz have already been fixed by the master
system. Demanding their mutual compatibility fixes three coupling constants
algebraically,
\begin{align}
\xi_1&=-\frac{\xi}{2+b^2\xi}\,,\qquad
\gamma_4=\frac{\xi-(2+3b^2\xi)\gamma_2}{b^2(4+5b^2\xi)}\,,
\nonumber\\
\gamma_1&=
\frac{-4\xi-2b^2\big[8\gamma_3+\xi(-\gamma_2+2\xi)\big]
+2b^4\xi(-16\gamma_3+\gamma_2\xi)-15b^6\gamma_3\xi^2}
{(2+b^2\xi)(4+5b^2\xi)}.
\label{eq:gamma1_constraint}
\end{align}
These relations are obtained by substituting the reduced equations back into
the remaining Einstein and bumblebee equations and requiring that the residual
coefficients vanish independently for arbitrary $Z(\phi)$, $V(\phi)$, and
charge parameters. In deriving these relations we assume
\begin{equation}
b\neq0,\qquad 1+b^2\xi\neq0,\qquad
2+b^2\xi\neq0,\qquad 4+5b^2\xi\neq0 .
\label{eq:nondegenerate_conditions}
\end{equation}
The degenerate cases require a separate analysis and will not be considered
here.
Thus we see that once the constraints on the couplings in
\eqref{eq:gamma1_constraint} are imposed, the complete set of equations of
motion closes on the three equations \eqref{eq:phi_master},
\eqref{eq:a_first_integral_hf} and \eqref{eq:f_master_general}. In particular, we note that \eqref{eq:f_master_general} already makes the asymptotic
structure manifest. For asymptotically trivial matter profiles, namely for
$r\rightarrow \infty$, we have
\begin{equation}
r^2V(\phi)\rightarrow 0\,,
\qquad
Z(\phi)\rightarrow Z_\infty\neq 0\,,
\qquad
f'(r)\rightarrow 0\,.
\label{eq:asymptotic_conditions}
\end{equation}
Then the source terms on the right-hand side of
\eqref{eq:f_master_general} vanish at large radius, and one obtains
\begin{equation}
f(\infty)=\frac{1}{1+\ell}\,,
\label{eq:f_infty}
\end{equation}
where $\ell$ is given by \eqref{LVvac}. Thus the metric is generally
asymptotic to the Lorentz-violating vacuum in the form of \eqref{LVvac}.
Imposing $h(\infty)=1$, which fixes the normalization of the time coordinate,
the constant $c$ is determined to be $c=1+\ell$. Using $\ell=b^2\xi$, the charge-renormalization factors become
\begin{align}
\Delta_e &=\frac{2(1+\ell) \left[10b^6\gamma_3\xi+b^4(8\gamma_3-3\gamma_2\xi)
-2b^2(\gamma_2-3\xi)+4\right]}{(2+\ell)(4+5\ell)}\,,\nn\\
\Delta_m &= -\frac{2(b^2\gamma_2-2)(1+\ell)}{4+5\ell}.
\label{eq:Delta_constrained}
\end{align}
The first-order differential equation of $f$ \eqref{eq:f_master_general} now becomes
\begin{equation}
r f'(r)+f(r)-\frac{1}{1+\ell}= -\frac{r^2}{2+\ell}\,V(\phi)
-\frac{4q^2\Delta_e}{r^2(1+\ell)\,Z(\phi)}
-\frac{4P_m^2\Delta_m\,Z(\phi)}{r^2(1+\ell)}\,.
\label{eq:f_master_constrained}
\end{equation}
In the next section, we shall consider specific NLED choices and construct
charged solutions in the bumblebee-deformed Einstein-NLED theory. The
construction proceeds in a definite order: first, the auxiliary scalar
$\phi$ is determined algebraically from \eqref{eq:phi_master}; then the
electric potential $a(r)$ and the metric function $f(r)$ are obtained from
\eqref{eq:a_first_integral_hf} and \eqref{eq:f_master_constrained},
respectively.

Recall that our original theory introduced five new parameters. The requirement of consistency \eqref{eq:gamma1_constraint} leaves two free parameters $(\gamma_2,\gamma_3)$, with $(\xi_1,\gamma_1,\gamma_4)$ solved in terms of these two and the bumblebee parameters $(\xi,b)$. If we turn on the magnetic charge as well, the radial consistency will impose a further condition, leaving only one free parameter. Since in this paper, we shall not consider magnetically charged solutions, we shall not present more details.

\section{Charged bumblebee-NLED geometries}
\label{sec:regular-bh}

In the previous section, we obtained the bumblebee-deformed Einstein-NLED
theory that admits spherically symmetric and static spacetime geometries
asymptotic to the Lorentz-violating vacuum \eqref{LVvac}.
We now construct explicit electrically charged solutions. The question of
whether these solutions describe black holes, marginally regular geometries,
or strictly regular horizonless geometries will be addressed by analyzing the
horizon condition and the near-origin expansion.
Although the reduced equations allow a magnetic parameter $P_m$, in the
explicit regularity analysis below we restrict to the purely electric branch
$P_m=0$. The genuinely dyonic case will be left for future work.

\subsection{An example: Born-Infeld theory}
\label{subsec:BI_example}

Born-Infeld theory is one of the best-known NLED theories in the
literature.
In four dimensions, the Lagrangian is
\begin{equation}
L=
\frac{2}{\alpha}\left(1-\sqrt{1+\alpha{\cal F}+\frac14\alpha^2\widetilde{\cal F}^{\,2}}
\right)\,,\qquad \widetilde{\cal F} =\frac18\epsilon^{\mu\nu\rho\sigma}
F_{\mu\nu}F_{\rho\sigma}.
\end{equation}
For simplicity, we focus on the purely electric branch with $P_m=0$, for
which $\widetilde{\cal F}=0$. This normalization gives $L_{\rm BI}=-{\cal F}+\mathcal O({\cal F}^2)$ in the weak-field limit.
The reduced Born-Infeld theory can then be
written as an MS theory \eqref{eq:EMS_auxiliary} with
\begin{equation}
Z(\phi)=\phi^{-1}\,, \qquad V(\phi) =\frac{1}{\alpha}\left(\phi+\phi^{-1}-2\right).
\label{eq:ZVforBI}
\end{equation}
Following the procedure described in the previous section, the electrically
charged solution can be constructed in closed form. We first obtain the
auxiliary scalar from \eqref{eq:phi_master}:
\begin{equation}
\phi_{\rm BI}(r)=\left(1+\frac{4\alpha q^2\Delta_e(2+\ell)} {(1+\ell)r^4}\right)^{-1/2}.
\label{eq:phi_BI_original_q}
\end{equation}
Choosing the gauge $a(\infty)=0$, the electric potential, integrated from
\eqref{eq:a_first_integral_hf}, is
\begin{equation}
a_{\rm BI}(r) = -\frac{4q\sqrt{1+\ell}}{r}
\,{}_2F_1\!\left( \frac14,\frac12;\frac54;
-\frac{4\alpha q^2\Delta_e(2+\ell)} {(1+\ell)r^4}
\right)\,.
\label{eq:a_BI_original_q}
\end{equation}
The metric function follows from the radial master equation
\eqref{eq:f_master_constrained} and is given by
\begin{equation}
f_{\rm BI}(r)= \frac{1}{1+\ell} -\frac{2m}{r} + \frac{2r^2}{3\alpha(2+\ell)}
\left[1- {}_2F_1\!\left(-\frac34,-\frac12;\frac14;-\frac{4\alpha q^2\Delta_e(2+\ell)}
{(1+\ell)r^4} \right)\right].
\label{eq:f_BI_original_q}
\end{equation}
This solution contains two integration constants, $m$ and $q$. The
parameter $m$ controls the Schwarzschild-type term, while $q$ is the electric integration parameter. The physical electric charge is proportional to $q\Delta_e$, as discussed below.

\subsection{A general class of NLED}
\label{subsec:NLED_general}

The MS theory with \eqref{eq:ZVforBI} is a special case of a
one-parameter family of auxiliary Maxwell-scalar theories \cite{Li:2024rbw}:
\begin{equation}
Z(\phi)=\phi^{-1}, \qquad V(\phi)=\frac{1}{\alpha\nu} \Bigl[
1+\nu\phi-(1+\nu)\phi^{\frac{\nu}{1+\nu}}\Bigr].
\label{eq:ZV_choice_EMS_general}
\end{equation}
The Born-Infeld example discussed above is recovered from this family by the analytically continued value $\nu=-1/2$.
Again, we consider the purely electric configuration with $P_m=0$.
Following the same construction, we obtain
\bea
\phi(r) &=& \left(1+ \frac{4\alpha q^2\Delta_e(2+\ell)}{(1+\ell)r^4}\right)^{-1-\nu}\,,\nn\\
a(r) &=& -\frac{4q\sqrt{1+\ell}}{r}
\,{}_2F_1\!\left( \frac14,1+\nu;\frac54;
-\frac{4\alpha q^2\Delta_e(2+\ell)} {(1+\ell)r^4}
\right)\,,
\label{eq:phi_EMS_general_sol_original_q}
\eea
together with the metric function
\begin{equation}
f(r) =\frac{1}{1+\ell}-\frac{2m}{r}+\frac{r^2}{3\alpha\nu(2+\ell)}\left[
{}_2F_1\!\left(-\frac34,\nu;\frac14;
-\frac{4\alpha q^2\Delta_e(2+\ell)}{(1+\ell)r^4}\right)-1\right].
\label{eq:f_EMS_general_sol_original_q}
\end{equation}
It is clear that when $\nu=-\frac12$,
the solution reduces to the BI case. The generic branch analyzed later will be taken with $\nu>0$, while the
Born-Infeld branch corresponds to the separate analytically continued value
$\nu=-1/2$.

Other auxiliary Maxwell-scalar choices can be bumblebee-deformed in the same
way, leading to additional electrically charged bumblebee-NLED solutions.
Since the construction follows directly from the three master equations
\eqref{eq:phi_master}, \eqref{eq:a_first_integral_hf}, and
\eqref{eq:f_master_constrained}, we shall not enumerate these further
examples here.

\subsection{Black hole thermodynamics}
\label{subsec:EMS_general_thermo}

The general electrically charged bumblebee-NLED solutions constructed in
the previous subsection contain two integration constants $(m,q)$. Before
imposing any center-regularity condition, these solutions describe black holes
whenever the metric function has a largest positive root. In this subsection,
we derive the thermodynamic quantities of this unconstrained black hole family, keeping the theory parameters $\alpha,\nu,\ell$, and $\Delta_e$ fixed. In the asymptotic region
$r\rightarrow\infty$, one finds
\be
-g_{tt}=h=(1 + \ell) f = 1 -\frac{2(1+\ell) m}{r} + \frac{4q^2\Delta_e}{r^2}
+\mathcal O(r^{-6})\,.
\label{eq:h_asymptotic_thermo}
\ee
This suggests that the mass should be proportional to the parameter $m$.
However, the precise proportionality can be tricky in a nonminimally coupled theory with the Lorentz-violating vacuum. Nevertheless, an event horizon can form for sufficiently large $m$. It is located at the largest root of $g_{tt}$, i.e.~$f(r_0)=0$. This condition relates $m$ to the horizon radius $r_0$ as
\begin{equation}
m = \frac{r_0}{2(1+\ell)}+ \frac{r_0^3}{6\alpha\nu(2+\ell)}
\left[
{}_2F_1\!\left(
-\frac34,\nu;\frac14;
-\frac{4\alpha q^2\Delta_e(2+\ell)} {(1+\ell)r_0^4}
\right)-1
\right].
\label{eq:m_EMS_general_horizon_original_q}
\end{equation}
Following the standard technique, we can determine the Hawking temperature, given by
\begin{equation}
T=\frac{\sqrt{1+\ell}}{4\pi}\left[\frac{1}{(1+\ell)r_0}+
\frac{r_0}{\alpha\nu(2+\ell)}
\left(
\left(1+\frac{4\alpha q^2\Delta_e(2+\ell)}
{(1+\ell)r_0^4}\right)^{-\nu}-1
\right)\right].
\label{eq:T_EMS_general_compact_original_q}
\end{equation}

For nonminimally coupled gravity theories, the natural starting point for the
black hole entropy is the Wald-Iyer Noether-charge formalism
\cite{Wald:1993,IyerWald:1994}. Similar subtleties in applying the Wald
formalism to nonminimally coupled matter sectors are known, for example, in
Einstein-Horndeski black holes with derivative coupling to the Einstein tensor
\cite{Feng:2015oea,Feng:2015wvb}. In the present bumblebee theory, a direct
horizon Noether-charge evaluation of the curvature coupling
$\xi B^\mu B^\nu R_{\mu\nu}$ gives
\be
S_{\rm W} = \pi (1 + \frac{1}{2}\ell ) r_0^2\,.
\label{eq:naive_Wald_entropy}
\ee
However, it turns out that this entropy formula is not consistent with the first law. The reason is the following: When the bumblebee vector is taken to be norm-fixed, the vector field is not an independent field variable on the horizon. The constraint $B^2=b^2$ ties the radial component $B_r=b/\sqrt{f}$ to the metric function, so that its variation is
correlated with the metric variation. Therefore, a direct evaluation of only
the explicit curvature derivative in the Noether charge does not by itself
give the entropy entering the integrable first law. A careful analysis of the complete constrained covariant phase-space variation gives instead \cite{Chen:2025ypx,Li:2025tcd}
\be
S= \pi \left(1+\ell\right) r_0^2\, .
\label{eq:correctS}
\ee
The corresponding conserved mass is then \cite{Chen:2025ypx,Li:2025tcd}
\be
M=(1+ \ell)^{\fft32} m\,.
\label{eq:mass_thermo}
\ee
Note that the coefficient before the parameter $m$ can be determined by turning off the electric charge. The electric charge contribution should not alter this relation.

We now turn our attention to the thermodynamic pair $(Q_e,\Phi_e)$.
The electric charge follows straightforwardly from the Maxwell equation \eqref{eq:A_general} and is given by
\begin{equation}
Q_e=\frac{1}{16\pi}\int_{S^2_\infty} {* {\cal P}_{(2)}} =q\Delta_e\,.
\label{eq:Qe_EMS_general_flux_original_q}
\end{equation}
With the gauge choice $a(\infty)=0$, the electric potential conjugate to
$Q_e$ is
\be
\Phi_e=a(\infty)-a(r_0) = \frac{4q\sqrt{1+\ell}}{r_0}\,
{}_2F_1\!\left(
\frac14,1+\nu;\frac54;
-\frac{4\alpha q^2\Delta_e(2+\ell)}{(1+\ell)r_0^4}
\right).
\label{eq:Phi_EMS_general_original_q}
\ee
It is now straightforward to verify the black hole thermodynamic first law holds, namely
\be
dM = T dS + \Phi_e dQ_e\,.
\label{firstlaw}
\ee
Extremal black holes, when they exist, are obtained by imposing
$T=0$, which relates $q$, $r_0$, and the fixed theory parameters through
\eqref{eq:T_EMS_general_compact_original_q}.

\section{Central regularity and horizon structure}
\label{sec:central_regular_horizon}

In the previous section, we constructed charged static and spherically symmetric solutions in bumblebee-deformed Einstein-NLED gravity. We now analyze their central behavior and horizon structure. The general NLED theory with $\nu>0$ was specifically constructed so that its minimal coupling to Einstein gravity admits electrically charged regular black holes when the mass and charge are fine-tuned to some appropriate ratio \cite{Li:2024rbw}. In addition to regular black holes, regular horizonless spacetimes can also emerge. It is therefore natural to ask whether regular solutions can also emerge in our bumblebee-deformed NLED gravity.

As pointed out above, the Lorentz-violating bumblebee geometries contain two distinct sources of singular behavior. The first comes from the conical Lorentz-violating background itself. For $\ell\neq0$, the areal-radius origin contains a conical curvature defect, with the Ricci scalar behaving as $R\sim 1/r^2$. The second is the usual black hole singularity associated with the Schwarzschild-type mass term, for which the Kretschmann scalar contains a stronger divergence of order $m^2/r^6$ as $r\rightarrow0$.  It is clear that the background $1/r^2$ singularity becomes insignificant when a black hole is present. However, it requires more stringent conditions to remove both types of singularities.

\subsection{Marginally regular black hole geometries}
\label{subsec:regular_marginal}

We first consider the generic case $\nu>0$, which in ordinary Einstein-MS gravity can give regular electric black holes after imposing an appropriate mass-charge relation. In the present bumblebee-deformed theory, the near-origin expansion of the metric function is
\begin{equation}
f(r)=-\frac{2(m-m_c)}{r} +\frac{1}{1+\ell} -
\frac{r^2}{3\alpha\nu(2+\ell)} + \mathcal O(r^{2+4\nu}),
\label{eq:f_EMS_general_near_origin_original_q}
\end{equation}
where $m_c$ is the critical mass determined by the electric charge parameter $q$, namely
\begin{equation}
m_c(q)=\frac{\Gamma\!\left(\frac14\right)
\Gamma\!\left(\nu+\frac34\right)}{6\alpha\nu\,\Gamma(\nu)(2+\ell)}
\left[\frac{4\alpha q^2\Delta_e(2+\ell)}
{1+\ell}\right]^{3/4}.
\label{eq:mc_EMS_general_original_q}
\end{equation}
Thus, the Schwarzschild-type pole can be removed by imposing
\begin{equation}
m=m_c .
\label{eq:m_EMS_general_star_original_q}
\end{equation}
Under this mass/charge relation, we have
\begin{equation}
f(r)=\frac{1}{1+\ell}-\frac{r^2}{3\alpha\nu(2+\ell)}+
\mathcal O(r^{2+4\nu}).
\label{eq:f_EMS_general_nonregular_original_q}
\end{equation}

Thus, in the asymptotically Minkowski limit $\ell=0$, the pole-removal condition $m=m_c$ gives the regular electric black hole solutions of the corresponding Einstein-NLED theory \cite{Li:2024rbw}. In the present Lorentz-violating background, however, $\ell\neq0$, and the constant term in the near-origin expansion remains
\begin{equation}
f(0)=\frac{1}{1+\ell}\neq1 .
\end{equation}
Therefore the Schwarzschild-type $1/r$ pole is removed, but the conical central defect of the Lorentz-violating vacuum persists. The singularity is weaker than the original Schwarzschild-type curvature singularity, but the center is still not regular in the areal-radius sense. We shall therefore refer to these pole-free configurations as marginally regular black hole geometries when an event horizon is present. Analogous spacetime can be found in charged black hole in Einstein-BI gravity when the mass/charge is taken to be a critical relation \cite{Li:2016nll}.

\begin{figure}[t]
\centering
\includegraphics[width=0.82\textwidth]{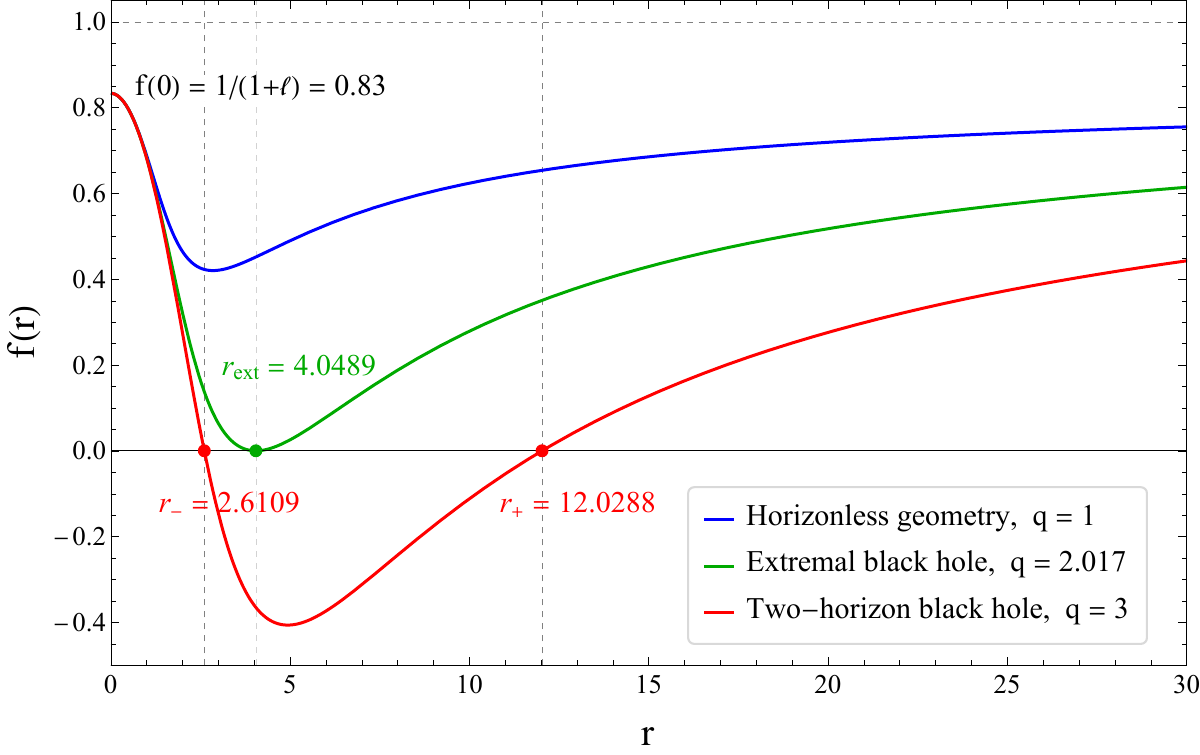}
\caption{\small
Metric function $f(r)$ for three representative pole-free geometries in the
generic $\nu>0$ branch. The common parameters are
$\alpha=1$, $\nu=1$, $b=1$, $\xi=1/5$, and $\ell=0.2$. The mass is fixed by the pole-removal condition
$m=m_c$. The dashed horizontal line marks
$f(0)=1/(1+\ell)\simeq0.83$, showing the residual conical central defect.
The blue curve with $q=1$ remains positive and is horizonless. The green
curve corresponds to the extremal value $q_{\rm ext}\simeq2.017$, with a
double horizon at $r_{\rm ext}\simeq4.0489$. The red curve with $q=3$ has
two horizons, $r_-\simeq2.6109$ and $r_+\simeq12.0288$.
}
\label{fig:marginal_regular_three_cases}
\end{figure}

After imposing $m=m_c$ so that the $1/r$ pole is removed, three qualitatively different horizon structures may occur: a marginally regular horizonless geometry, an extremal
marginally regular black hole, or a nonextremal marginally regular black hole,
depending on whether $f(r)$ has no positive root, a double positive root, or
two positive roots, respectively. For fixed theory parameters, these
possibilities are controlled by the electric parameter $q$, as illustrated in
Fig.~\ref{fig:marginal_regular_three_cases}.

\subsection{Regular horizonless geometries}
\label{subsec:regular_horizonless}

In the previous subsection, the mass parameter was tuned to remove the Schwarzschild-type $1/r$ pole. However, for $\ell\neq0$, the pole-free generic $\nu>0$ branch still has $f(0)=1/(1+\ell)\neq1$, and therefore retains a residual conical central defect. A regular center requires two independent conditions: the $1/r$ pole must vanish, and the constant term in $f(r)$ must be adjusted to unity. This is why the Born-Infeld branch is special: the electric parameter contributes to the constant term in the near-origin expansion.

The situation changes when $\nu=-1/2$, corresponding precisely to the BI theory, and the bumblebee-deformed black hole is given by \eqref{eq:f_BI_original_q}. The reality condition for $r\in [0,\infty)$ requires
\begin{equation}
\frac{4\alpha q^2\Delta_e(2+\ell)}{1+\ell}>0\,.
\label{eq:BI_real_branch_condition}
\end{equation}

Since $\Delta_e$ is fixed by the nonminimal bumblebee-electromagnetic couplings, the reality condition can be satisfied for either sign of $\alpha$, provided that $\Delta_e$ is chosen such that $\alpha\Delta_e>0$. The near-origin expansion of \eqref{eq:f_BI_original_q} is
\begin{equation}
f_{\rm BI}(r) =-\frac{2(m-m_{\rm BI})}{r} +\left[\frac{1}{1+\ell}- \frac{2|q|}{\alpha(2+\ell)} \sqrt{\frac{4\alpha\Delta_e(2+\ell)} {1+\ell}}
\right] + \frac{2r^2}{3\alpha(2+\ell)}+\mathcal O(r^4),
\label{eq:f_BI_near_origin}
\end{equation}
where
\begin{equation}
m_{\rm BI}(q) =\frac{\Gamma\!\left(\frac14\right)^2}{6\sqrt{\pi}\,\alpha(2+\ell)}
\left[\frac{4\alpha q^2\Delta_e(2+\ell)}{1+\ell}\right]^{3/4}.
\label{eq:m_BI_regular_mass}
\end{equation}

Unlike the generic $\nu>0$ case, the Born-Infeld branch contains an electric-charge-dependent constant term in the near-origin expansion, as shown in the square bracket of \eqref{eq:f_BI_near_origin}. This feature is absent in the $\nu>0$ case and is precisely what allows the constant term to be tuned to unity. In ordinary Einstein-Born-Infeld gravity, corresponding to $\ell=0$, this tuning is not available: the electric field is regular at the origin, but the metric still retains a curvature singularity. In the bumblebee-deformed case, however, the conical background contribution $1/(1+\ell)$ can be compensated by the Born-Infeld charge contribution. Together with the removal of the Schwarzschild-type pole, we find that both the mass and electric charge are completely determined as
\begin{equation}
m=m_{\rm BI}, \qquad |q|= -\frac{\alpha\ell}{4} \sqrt{
\frac{2+\ell}{\alpha\Delta_e(1+\ell)}}\, .
\label{eq:mq_BI_tuning}
\end{equation}
The resulting $f$ near the origin becomes
\begin{equation}
f_{\rm BI}(r) = 1+ \frac{2r^2}{3\alpha(2+\ell)} +\mathcal O(r^4)\,.
\label{eq:f_BI_regular_center}
\end{equation}
Since $1+\ell>0$, one also has $2+\ell>0$. The sign of the
quadratic term in \eqref{eq:f_BI_regular_center} is therefore controlled by
$\alpha$. Thus $\alpha>0$ gives an AdS core, whereas $\alpha<0$ gives
a dS core. For $|q|$ in \eqref{eq:mq_BI_tuning} to be real and
non-negative, the regular branch must satisfy
\[
\alpha\Delta_e>0,\qquad \alpha\ell<0 .
\]
Since $\Delta_e$ depends on the remaining nonminimal electromagnetic
couplings, its sign can be chosen consistently with the sign of $\alpha$.
The real regular Born-Infeld branch therefore splits into two cases:
\begin{table}[H]
\centering
\renewcommand{\arraystretch}{1.25}
\caption{Two branches of regular Born-Infeld geometries.}
\label{tab:BI_regular_two_branches}
\begin{tabular}{ccc}
\hline\hline
\textbf{Core type} & \textbf{Conditions} & \textbf{Properties} \\
\hline
AdS core &
$\alpha>0,\ \Delta_e>0,\ -1<\ell<0$ & $m_{\rm BI}>0,\ f(\infty)>1$ \\
dS core &
$\alpha<0,\ \Delta_e<0,\ \ell>0$ & $m_{\rm BI}<0,\ f(\infty)<1$ \\
\hline\hline
\end{tabular}
\end{table}

\begin{figure}[t]
\centering
\includegraphics[width=0.82\textwidth]{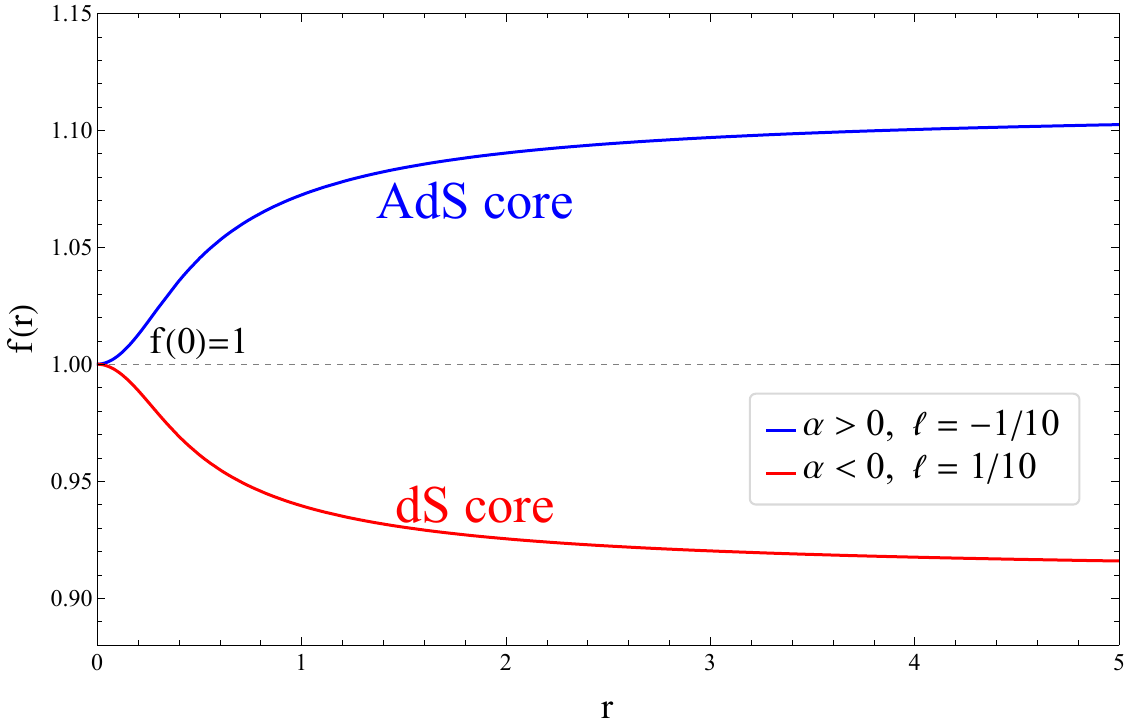}
\caption{\small
Near-origin behavior of the regular Born-Infeld branch. The blue curve
represents an $\alpha>0$ branch with $\ell=-1/10$, which has an AdS core.
The red curve represents an $\alpha<0$ branch with $\ell=1/10$, which has
a dS core. In both cases the regularity conditions have been imposed so that
$f(0)=1$, and the Schwarzschild-type $1/r$ pole is absent.
}
\label{fig:BI_regular_core}
\end{figure}

Thus the strictly regular real Born-Infeld branch naturally splits into an $\alpha>0$ branch and an $\alpha<0$ branch. The behavior of $f(r)$ from the origin to the asymptotic region is shown in Fig.~\ref{fig:BI_regular_core}.
We see that both solutions with AdS or dS cores are completely regular, describing horizonless geometries. Since both the mass and electric charge are used to fine-tune regularity,
there is no remaining free parameter with which to form a regular black hole.

It is useful to distinguish the regular branch from nearby singular
black hole branches. If $q$ is fixed by \eqref{eq:mq_BI_tuning} but $m\neq m_{\rm BI}$, the $1/r$ pole in \eqref{eq:f_BI_near_origin} reappears. In particular, for $m>m_{\rm BI}$, one has $f_{\rm BI}(r)\rightarrow-\infty$ as $r\rightarrow0$, while $f_{\rm BI}(\infty)=1/(1+\ell)>0$. Therefore at least one positive horizon root exists. Such solutions are black holes, but their centers are singular.
They should not be interpreted as regular black holes. Thus, within the real purely electric Born-Infeld branch analyzed here, imposing central regularity yields a horizonless geometry rather than a regular black hole.

\FloatBarrier

\section{Conclusion}
\label{sec:con}

In this work, we constructed electrically charged Lorentz-violating geometries in bumblebee gravity nonminimally coupled to a general class of NLED theories written in the auxiliary Maxwell-scalar form. We focused on the norm-fixed radial bumblebee vector, for which the asymptotic geometry is a conical Lorentz-violating vacuum. The nonminimal couplings between the bumblebee vector $B_\1$ and the NLED sector are stringently required by the consistency of the equations of motion.

We began with a general class of bumblebee-NLED couplings up to and including quartic order in the bumblebee vector. This introduces five new parameters, in addition to the original bumblebee parameters $(\xi,b)$. Requiring consistency of the full static and spherically symmetric field equations fixes three of these parameters algebraically. After this reduction, the system is governed by a closed set of radial differential equations for the auxiliary scalar, the electric potential, and the metric function $-g_{tt}\sim 1/g_{rr}$. This provides a systematic way to construct charged bumblebee-NLED solutions.

We then applied this framework to Born-Infeld theory and to a one-parameter family of NLED theories that includes Born-Infeld theory as a special case. Exact solutions were obtained for all these theories, involving two integration constants, parametrizing the mass and electric charges $(M,Q)$. For sufficiently large mass, these solutions describe electrically charged black holes that are asymptotic to a Lorentz-violating spacetime given by the direct product of time and a three-dimensional Euclidean cone. We computed the thermodynamic quantities and verified the first law of black hole thermodynamics.

We analyzed the global structure, focusing on the possibility of constructing regular spacetimes. The general solutions with $(M,Q)$ have two sources of curvature singularity at the center $r=0$: the Schwarzschild-type pole and the residual conical singularity of the Lorentz-violating vacuum itself. In previously considered Einstein-NLED gravity, suitable NLED theories can be constructed so that regular black holes arise after fine-tuning the mass-charge relation to remove the Schwarzschild-type pole. In the present bumblebee-NLED gravity, the Schwarzschild-type pole can also be removed, but the conical central defect remains. This gives rise to marginally regular black holes.

In Einstein-Born-Infeld gravity, the charged black hole can be fine-tuned to remove the Schwarzschild-type pole singularity, but the geometry remains only marginally regular because a conical singularity persists. However, in the present bumblebee-deformed construction, by contrast, this conical singularity can be further removed by the residual contribution at $r=0$ induced by the Lorentz-violating asymptotic structure. Completely regular spacetimes can therefore be constructed in bumblebee-deformed Born-Infeld gravity. However, this requires both the mass and the charge to be fixed entirely by the coupling constants. Moreover, the regular solutions are not black holes, but horizonless geometries interpolating from AdS or dS cores to Lorentz-violating vacua. Intriguingly, the mass of the latter smooth geometry requires negative mass.

Thus, in addition to the subtleties involved in finding a consistent bumblebee-deformed matter system,
the existence of the conical singularity in the Lorentz-violating vacuum itself makes the construction of regular black holes in bumblebee-deformed NLED gravity much more complicated than in Einstein-NLED gravity. A simple counting of parameters indicates that regular black holes would require an additional NLED matter sector to provide an extra free parameter for satisfying both the regularity and horizon conditions.

\section*{Acknowledgements}

This work is supported in part by the National Natural Science Foundation of China (NSFC) grants No.~12375052 and No.~11935009, and also by the Tianjin University Self-Innovation Fund Extreme Basic Research Project Grant No.~2025XJ21-0007.


\end{document}